\documentclass[aps,amssymb,amsmath, twocolumn, pra, superscriptaddress]{revtex4-1}
\usepackage{graphicx}
\usepackage{epsfig}
\usepackage{amsmath}
\usepackage{bm}
\usepackage{braket}
\usepackage{natbib}
\usepackage{enumitem}
\usepackage{longtable}
\usepackage{natbib}
\newcommand{\BH}{\mathcal{B}(\mathcal{H})}
\def\oper{{\mathchoice{\rm 1\mskip-4mu l}{\rm 1\mskip-4mu l}%
{\rm 1\mskip-4.5mu l}{\rm 1\mskip-5mu l}}}

\usepackage[colorlinks, citecolor= magenta]{hyperref}
\usepackage[up]{subfigure}

\usepackage{epstopdf}

\usepackage{booktabs}

\usepackage{amsthm}
\newtheorem{prop}{Proposition}
\newtheorem{thm}{Theorem}

\newtheorem{ex}{Example}

\newtheorem{cor}{Corollary}

\begin{document}

\title{Information flow versus divisibility for qubit evolution}
\author{Sagnik Chakraborty}
\email{csagnik@imsc.res.in}
\affiliation{Optics and Quantum Information Group, The Institute of Mathematical
Sciences, C. I. T. Campus, Taramani, Chennai 600113, India}
\affiliation{Homi Bhabha National Institute, Training School Complex, Anushakti Nagar, Mumbai 400094, India}

\author{Dariusz Chru\'sci\'nski}
\email{darch@fizyka.umk.pl}
\affiliation{Institute of Physics, Faculty of Physics, Astronomy and Informatics, Nicolaus Copernicus University,
Grudziądzka 5/7, 87–100 Toru\'n, Poland}

\begin{abstract}
{We {study the relation} between lack of Information Backflow and completely positive divisibility (CP divisibility) for non-invertible qubit dynamical maps.  Recently, { these two concepts were shown to be fully equivalent} for the so called  image non-increasing dynamical maps. Here we show that this equivalence  is universal for any qubit dynamical map. A key ingredient in our proof is the observation that there does not exist CPTP projector onto a 3-dimensional subspace spanned by qubit density operators. Our analysis is illustrated by several examples of qubit evolution including also  dynamical maps which are not image non-increasing. % to emphasize the importance of our result.
}
\end{abstract}
%\pacs{03.65.Yz, 03.65.Ta, 42.50.Lc}

\maketitle
\section{Introduction}

%\noindent {\em Introduction.} ---
With the discovery of new experimental techniques, Quantum Information is no longer a subject of solely theoretical interest. Numerous quantum protocols, principles and results discovered theoretically, are now being tested, verified and reinforced upon, by experiments. Along with these new developments, comes a serious challenge of controlling and understanding real life quantum systems, which are inherently open to the environment. This has resulted in drawing a lot of interest about open quantum systems in recent times \cite{open1,open2,open3,NM1,exp,exp2,exp3}. The evolution of such systems is represented by a dynamical map, that is, a family of completely positive (CP) trace-preserving (TP) maps $\Lambda_t : \BH \to \BH$ ($t \geq 0$), where $\BH$ denotes a linear space of bounded operators acting on the system Hilbert space $\mathcal{H}$ (actually, in this paper we consider only finite dimensional case and hence $\BH$ coincides with all linear operators on $\mathcal{H}$). Moreover, one assumes a natural initial condition $\Lambda_{t=0} = {\rm id}$ (identity map).

Usually, the origin of a dynamical map is a composed system living in $\mathcal{H} \otimes \mathcal{H}_E$, with $\mathcal{H}_E$ denoting a Hilbert space of environment. Now, if $\mathbf{H} = H_S + H_E + H_{\rm int}$ is the total Hamiltonian of the composed system and $\rho\otimes \rho_E$ is an initial product state, then the standard reduction procedure defined via partial trace operation

\begin{equation}\label{}
  \Lambda_t(\rho) = {\rm Tr}_E \Big( e^{-i \mathbf{H} t}\, \rho \otimes \rho_E \, e^{i \mathbf{H} t} \Big) ,
\end{equation}
gives rise to a legitimate dynamical map (in the paper we keep $\hbar=1$).

Recently, the notion of non-Markovian quantum evolution has received considerable attention (see review papers \cite{open1,open2,open3, NM1}). This property, although well defined in the classical regime \cite{NM1,NM2,NM3,NM4}, has a number of non-equivalent prescriptions in quantum theory. On the level of dynamical maps two main approaches which turned out to be
very influential are based on the concept of CP divisibility \cite{RHP} and information flow \cite{BLP}. One calls a dynamical map $\Lambda_t$ divisible if

\begin{equation}
 \label{cpd_defn}
\Lambda_t=V_{t,s}\circ \Lambda_s \ \ \ , \ (t \geq s) ,
\end{equation}
where $V_{t,s} : \BH \to \BH$ is a linear map defined on the entire $\BH$. Being divisible the map $\Lambda_t$ is:

 $i)$  P-divisible if the map $V_{t,s}$ is positive and trace-preserving (PTP), and

 $ii)$ CP-divisible if the map $V_{t,s}$ is CPTP \cite{footnote1}.

(For a mathematical details of  positive and completely positive maps see \cite{Paulsen, Erling}).

In the latter case one may interpret $V_{t,s}$ as a legitimate quantum channel, mapping states at time $s$ into states at time $t$. Following \cite{RHP} one calls the quantum evolution to be Markovian iff the corresponding dynamical map is CP-divisible.

A second idea developed in  \cite{BLP} is based upon the notion of information flow: for any pair of density operators $\rho_1$ and $\rho_2$ one defines an information flow
\begin{equation}\label{BLP}
  \sigma(\rho_1,\rho_2;t) = \frac{d}{dt} \|\Lambda_t(\rho_1) - \Lambda_t (\rho_2)\|_1 ,
\end{equation}
where $\|A\|_1= {\rm Tr}\sqrt{AA^\dagger}$ denotes the trace norm of $A$. Actually $\|\rho_1-\rho_2\|_1$ represents distinguishability of $\rho_1$ and $\rho_2$. Moreover, $\frac 12(1+ \|\rho_1-\rho_2\|_1)$ gives the maximal guessing probability in the biased scenario, that is,  when $\rho_1$ and $\rho_2$ are prepared with the same probability \cite{NC00}. Following \cite{BLP} Markovian evolution is characterized by the condition $\sigma(\rho_1,\rho_2;t) \leq 0$ for any pair of initial states $\rho_1$ and $\rho_2$. Whenever $\sigma(\rho_1,\rho_2;t) > 0$ one calls it information backflow meaning that the information flows from the environment back to the system. In this case the evolution displays nontrivial memory effects and it is evidently non-Markovian. One calls $ \sigma(\rho_1,\rho_2;t) \leq 0$ a BLP condition.

In this paper, we address the problem of analyzing how far these two approaches to Markovianity are equivalent. It was shown in \cite{Angel} that for invertible dynamical maps CP-divisibility is equivalent to lack of information backflow on an extended system comprised of the system and an $d$-dimensional ancilla \cite{BOGNA} in an extended scenario when two states $\rho_1$ and $\rho_2$ are prepared with probabilities $p_1$ and $p_2$. Authors of \cite{BOGNA} observed that one may still restrict to the biased case $p_1=p_2$ but the price one pays is the use of $(d+1)$-dimensional ancilla.  These results were then extended to image non-increasing  dynamical maps \cite{PRL-2018}, which is a large class of dynamical maps including all invertible ones. Also recently the equivalence between divisibility and monotonic decrease of information in terms of guessing probability was studied by Buscemi and Datta  in \cite{datta} for time discrete dynamical maps.
We show here that the result of \cite{PRL-2018} can be extended to arbitrary dynamical map if we restrict to dynamical maps on qubits. Our results proves complete equivalence of the two main approaches to Markovianity for qubit dynamical maps.

%A key ingredient of our proof is a obsevat that there can be no CPTP projector onto a 3-dimensional subspace spanned by qubit states.

The paper is structured as follows: in Section \ref{Background}, we review the recent results in this direction so as to provide a background   for the paper. Next in Section \ref{equivalence}, we present the main result of our paper and in Sectin \ref{examples}, we discuss some examples before drawing our conclusions in Section \ref{Conclusion}.

%\vspace{.4cm}
\section{Invertible vs. non-invertible maps}%Background}
\label{Background}
%\noindent {\em Invertible maps.} ---

By invertible dynamical map we understand $\Lambda_t$ such that $\Lambda_t^{-1}$ exists for all $t >0$. Note, that even if it exists the inverse need not be completely positive (it is always trace-preserving). The inverse is also completely positive if and only if the map $\Lambda_t$ is unitary, that is, $\Lambda_t(\rho) = U_t \rho U_t^\dagger$, where $U_t$ is time-dependent unitary operator in $\mathcal{H}$. Now, invertible map is always divisible. Indeed one finds $V_{t,s} = \Lambda_t \Lambda_s^{-1}$. Moreover,

\begin{thm}[\cite{Angel}] \label{TH-I} An invertible dynamical map $\Lambda_t$ is P-divisible if and only if

\begin{equation}\label{P}
  \frac{d}{dt} \|\Lambda_t(p_1\rho_1 - p_2\rho_2)\|_1 \leq 0 ,
\end{equation}
for all probability distributions $p_1+p_2=1$ and  density operators $\rho_1,\rho_2$ in $\mathcal{H}$. Moreover, it is CP-divisibe if and only if

\begin{equation}\label{CP}
  \frac{d}{dt} \|[{\rm id}_d \otimes \Lambda_t](p_1\varrho_1 - p_2\varrho_2)\|_1 \leq 0 ,
\end{equation}
for all probability distributions $p_1+p_2=1$ and  density operators $\varrho_1,\varrho_2$ in $\mathbb{C}^d \otimes \mathcal{H}$ (with $d = {\rm dim}\, \mathcal{H}$).
\end{thm}
Note, that BLP Markovianity condition coincides with (\ref{P}) with $p_1=p_2$. Interestingly, one may restrict to completely biased case ($p_1=p_2$) due to the following

\begin{thm}[\cite{BOGNA}]\label{TH-2} An invertible dynamical map $\Lambda_t$ is CP-divisibe if and only if
\begin{equation}\label{CP-Bogna}
  \frac{d}{dt} \|[{\rm id}_{d+1} \otimes \Lambda_t](p_1\varrho_1 - p_2\varrho_2)\|_1 \leq 0 ,
\end{equation}
for all density operators $\varrho_1,\varrho_2$ in $\mathbb{C}^{d+1} \otimes \mathcal{H}$.
\end{thm}

%\vspace{.3cm}

%\noindent {\em Non-invertible maps.} ---

For maps which are not invertible even divisibility is not evident \cite{BOGNA,PRL-2018}. Note, that $V_{t,s}$ is well defined on the image of the maps $\Lambda_s$ (we denote it by ${\rm Im}(\Lambda_s)$). Actually, as shown in \cite{PRL-2018} divisibility of $\Lambda_t$ is equivalent to the following property

\begin{equation}\label{}
  {\rm Ker}(\Lambda_s) \subseteq {\rm Ker}(\Lambda_t) ,
\end{equation}
{for any $s<t$}, that is, the map is {\em kernel non-decreasing}. This condition guarantees that $V_{t,s}$ can be consistently extended from ${\rm Im}(\Lambda_s)$ to the whole space $\BH$. Authors of \cite{PRL-2018} analyzed the following question: {\em when the extension of $V_{t,s}$ is CPTP?}
The central result of \cite{PRL-2018} states

\begin{thm}[\cite{PRL-2018}] If the  dynamical map $\Lambda_t$ satisfies

\begin{equation}\label{CP-not}
  \frac{d}{dt} \|[{\rm id}_d \otimes \Lambda_t](p_1\varrho_1 - p_2\varrho_2)\|_1 \leq 0 ,
\end{equation}
for all probability distributions $p_1+p_2=1$ and  density operators $\varrho_1,\varrho_2$ in $\mathbb{C}^d \otimes \mathcal{H}$, then it is divisible with $V_{t,s}$ completely positive on $\BH$ but trace-preserving only on the image ${\rm Im}(\Lambda_s)$.
\end{thm}
Interestingly, there exists a class of dynamical maps for which the extension of $V_{t,s}$ is not only completely positive but also trace-preserving, that is, such maps are CP-divisible. One call a dynamical map $\Lambda_t$ {\em image non-increasing} \cite{PRL-2018} if

\begin{equation}\label{}
  {\rm Im}(\Lambda_t) \subseteq {\rm Im}(\Lambda_s) ,
\end{equation}
for $t > s$.

\begin{thm}[\cite{PRL-2018}] \label{TH-III} If the  dynamical map $\Lambda_t$ is image non-increasing and it satisfies

\begin{equation}\label{CP-not}
  \frac{d}{dt} \|[{\rm id}_d \otimes \Lambda_t](p_1\varrho_1 - p_2\varrho_2)\|_1 \leq 0 ,
\end{equation}
for all probability distributions $p_1+p_2=1$ and  density operators $\varrho_1,\varrho_2$ in $\mathbb{C}^d \otimes \mathcal{H}$, then it is CP-divisible.
\end{thm}

%{\color{red}
Finally,  Theorem \ref{TH-2} may be generalized as follows

\begin{thm}[\cite{PRL-2018}] \label{TH-IV} If the  dynamical map $\Lambda_t$ is image non-increasing and it satisfies

\begin{equation}\label{CP-not}
  \frac{d}{dt} \|[{\rm id}_{d+1} \otimes \Lambda_t](\varrho_1 - \varrho_2)\|_1 \leq 0 ,
\end{equation}
for all
%probability distributions $p_1+p_2=1$ and
density operators $\varrho_1,\varrho_2$ in $\mathbb{C}^{d+1} \otimes \mathcal{H}$, then it is CP-divisible.
\end{thm}
%}

%\vspace{.3cm}
\section{%Equivalence between Information flow and divisibility for
Qubit dynamical maps}
\label{equivalence}
%\noindent {\em Non-invertible qubit maps.} ---

Now, we consider the simplest scenario -- dynamical maps for qubits. The main result of this paper is provided by the following

\begin{thm}\label{thm_main_new} A qubit  dynamical map $\Lambda_t$ is CP-divisible if and only if

\begin{equation}\label{CP-q}
  \frac{d}{dt} \|[{\rm id}_2 \otimes \Lambda_t](p_1\varrho_1 - p_2\varrho_2)\|_1 \leq 0 ,
\end{equation}
for all probability distributions $p_1+p_2=1$ and 2-qubit density operators $\varrho_1,\varrho_2$ in $\mathbb{C}^2 \otimes \mathbb{C}^2$.
\end{thm}
We stress that this results is universal, that is, we do not assume that the map is invertible (as in Theorem \ref{TH-I}) nor that it is image non-increasing (as in Theorem \ref{TH-III}). Of course for invertible qubit maps it is just special case of Theorem \ref{TH-I}.

The proof of this result is based on the following observations:
% \begin{prop}[\cite{QUBIT}] \label{PRO-qubit} If $\Phi : M_2(\mathbb{C}) \to M_2(\mathbb{C})$ is a quantum channel, then ${\rm dim}\, {\rm Im}(\Phi) \in \{1,2,4\}$, that is, there is no qubit channel with single vanishing eigenvalue.
% \end{prop}
It was shown in \cite{PRL-2018} that if $\Lambda_t$ is  not invertible and that $t_1>0$ is the first moment of time such that $\Lambda_{t_1}^{-1}$ does not exist then condition (\ref{CP-q}) implies that

\begin{equation}
 \Pi_{t_1}=\lim_{\epsilon\rightarrow0^+} V_{t_1,t_1-\epsilon} ,
\end{equation}
defines a completely positive projector  onto $\text{Im}(\Lambda_{t_1})$.
%{\color{red}Note that this result is valid even if $\Lambda_t$ is just divisibe \cite{PRL-2018}.} Interestingly, for qubit evolution one has the %following

\begin{prop}
\label{lem_subspace}
There is no CPTP projector $\Pi : M_2(\mathbb{C}) \to M_2(\mathbb{C})$ onto a 3-dimensional subspace of $M_2(\mathbb{C})$ spanned by density operators.
\end{prop}

%{\color{red}

Actually, the above Proposition follows from the following  result of \cite{QUBIT} (however, we provide an independent proof in Appendix \ref{proof_proposition1}): let $\Phi$ be a qubit quantum channel and let ${\rm PO}(\Phi)$ be the pure output of $\Phi$, that is, the set of pure state in the image of $\Phi$

$$     {\rm PO}(\Phi) = \Phi(\mathbf{B}) \cap \mathbf{S} , $$
where $\mathbf{B}$ is a Bloch ball and $\mathbf{S}$ a Bloch sphere --- a set of qubit pure states. One proves

\begin{prop}[\cite{QUBIT}] \label{QUBIT} Let $\Phi$ be a qubit quantum channel such that ${\rm PO}(\Phi)$ has more than two elements. Then ${\rm PO}(\Phi) = \mathbf{S}$, that is, all pure states belong to the pure output of $\Phi$.
\end{prop}
Now, suppose that there exists a CPTP projector $\Phi$ such that its image is 3-dimensional. {Being a projector, it does not change the purity of the input states in subspace ${\rm Im}(\Phi)$.} It is, therefore, clear that the intersection ${\rm Im}(\Phi) \cap \mathbf{S}$ defines a circle on the Bloch sphere. But it contradicts Proposition \ref{QUBIT} which requires that in this case ${\rm PO}(\Phi) = \mathbf{S}$.

%\begin{Remark}

\begin{figure*}
  \includegraphics[scale=0.6]{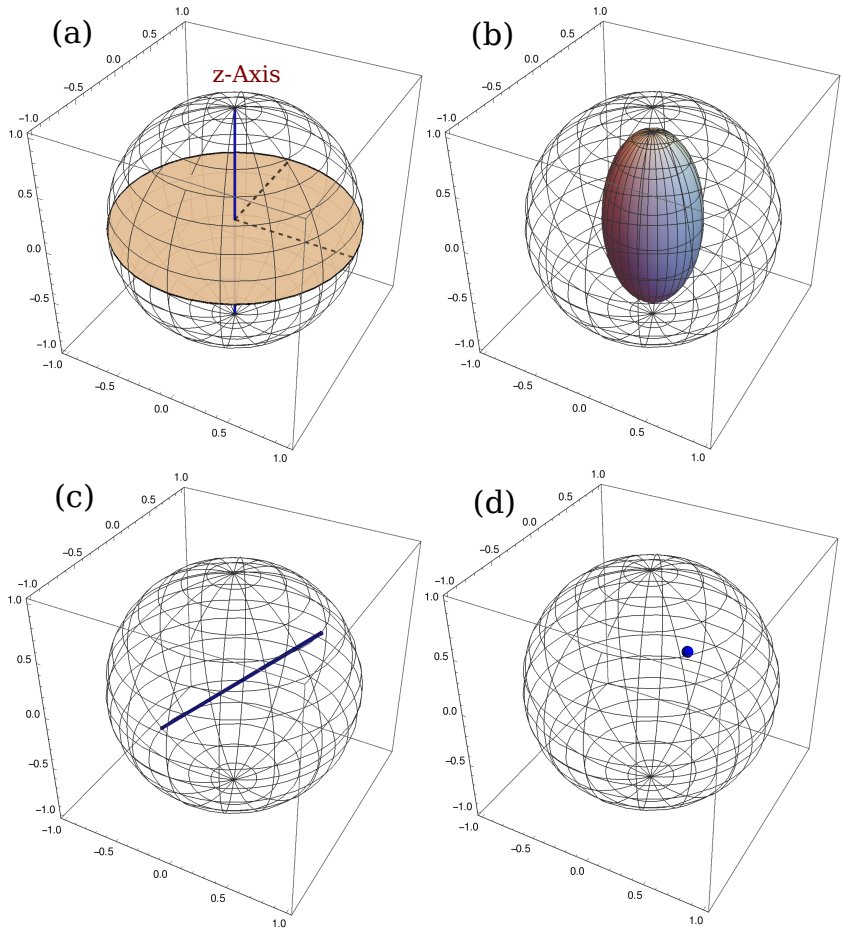}
  \caption{$(a)$ Bloch ball representation of the action of the PTP map $\Psi$ and the CPTP map $\mathcal{S}_3$. The equatorial brown disc and the thick blue z-Axis represent the image of the input bloch ball under the action of maps $\Psi$ and $\mathcal{S}_3$, respectively.
  $(b)$, $(c)$ and $(d)$ represents the allowed structures of the image of the bloch ball under action of a qubit dynamical map. $(b)$ When the map is invertible the image is an ellipsoid. $(c)$ When the map is non-invertible, and its image is 2 dimensional it forms a line within the bloch ball. $(d)$ When the map is non-invertible, and its the image has dimension 1 it forms a point.}
  \label{bloch}
 \end{figure*}

Note, that Proposition \ref{lem_subspace} does not forbid the existence of qubit quantum channel $\Phi$ such that ${\rm dim}\,{\rm Im}(\Phi)=3$. As an example consider

\begin{equation}\label{}
  \Phi(\rho) = \frac 12  \rho + \frac 14 \Big( \sigma_1 \rho \sigma_1 + \sigma_2 \rho \sigma_2 \Big) .
\end{equation}
One finds

\begin{equation*}\label{}
   \Phi(\oper) = \oper , \ \Phi(\sigma_1) = \frac 12 \sigma_1 , \ \Phi(\sigma_2) = \frac 12 \sigma_2 , \ \Phi(\sigma_3)= 0 ,
\end{equation*}
which proves that the range of $\Phi$ is 3-dimensional. It is clear that $\Phi$ being a CPTP map is not a CPTP projector (it has two eigenvalues $1/2$), {and hence does not preserve the purity of the input states in ${\rm Im}(\Phi)$}. It should  be stressed that this result is no longer true for positive and trace-preserving projectors. Consider a map

\begin{equation}\label{}
   \Psi(\rho) = \frac 14\Big( 3 \rho +  \sigma_1 \rho \sigma_1 + \sigma_2 \rho \sigma_2 - \sigma_3 \rho \sigma_3 \Big) .
\end{equation}
One finds

$$   \Psi(\oper) = \oper , \ \Psi(\sigma_1) =  \sigma_1 , \ \Psi(\sigma_2) =  \sigma_2 , \ \Psi(\sigma_3)= 0 , $$
and hence $\Psi$ maps a density operator

\begin{equation}\label{}
   \rho = \frac{1}{2}\Big( \oper + x_1 \sigma_1 + x_2 \sigma_2 + x_3 \sigma_3 \Big)
\end{equation}
to a density operator

\begin{equation}\label{}
   \Psi(\rho) = \frac{1}{2}\Big( \oper + x_1 \sigma_1 + x_2 \sigma_2\Big) ,
\end{equation}
that is, $\Psi$ projects a Bloch ball into a disk $x_3=0$. For more details cf.  Appendix \ref{positive-projector}. Interestingly a map projecting a Bloch ball to $x_3$ axis defined by

\begin{equation}\label{}
   \mathcal{S}_3(\rho) = \frac 12\Big(  \rho +  \sigma_3 \rho \sigma_3 \Big) ,
\end{equation}
is CPTP projector satisfying

$$   \mathcal{S}_3(\oper) = \oper , \ \mathcal{S}_3(\sigma_1) =  0 , \ \mathcal{S}_3(\sigma_2) =  0 , \ \mathcal{S}_3(\sigma_3)= \sigma_3 , $$
and hence ${\rm dim}\, {\rm Im}(\mathcal{S}_3)=2$. A pictorial representation of the action of $\Psi$ and $\mathcal{S}_3$ is given in Fig. \ref{bloch}$(a)$.

%  \begin{figure}
%   \includegraphics[width=\columnwidth]{qubit-dyn-bloch.eps}
%   \caption{Bloch ball representation of the action of the PTP map $\Psi$ and the CPTP map $\mathcal{S}_3$. The equatorial brown disc and the thick blue z-Axis represent the image of the input bloch ball under the action of $\Psi$ and $\mathcal{S}_3$, respectively.}
%   \label{qubit-dyn}
%  \end{figure}
%

% ======================
%
% {\bf Maybe we can add a figure representing CPTP projection to $x_3$ and PTP projection to the plane $x_3=0$. What do you think? Any idea for other figures?}
%
% =================

Now, observe that for the map $\Psi$ only points from the equator

$$ \frac{1}{2}\Big( \oper + \cos\phi \sigma_1 + \sin\phi \sigma_2\Big) ,\ \ \phi\in [0,2\pi) , $$
belong to ${\rm PO}(\Psi)$. Hence this  map cannot be completely positive. Actually, one proves

\begin{prop} Let $\Pi : {M}_2(\mathbb{C}) \to M_2(\mathbb{C})$  be a positive trace-preserving projector { onto a 3 -dimensional subspace}. Then ${\rm PO}(\Pi)$ is a great circle on the Bloch ball $\mathbf{S}$. Equivalently the subspace ${\rm Im}(\Phi) = \Phi(M_2(\mathbb{C}))$ is an operator system (contains $\oper$ and is closed under hermitian conjugation \cite{bhatia2009positive}).
\end{prop}
For the proof see Appendix \ref{positive-projector}. The above discussion lead us to the following

\begin{cor} If the qubit dynamical map $\Lambda_t$ satisfies (\ref{CP-q}), then the dimension of its image ${\rm dim}\,{\rm Im}(\Lambda_t) \in  \{1,2,4\}$.
\end{cor}

%\end{Remark}
%}

A pictorial representation of the above result is given in Figs. \ref{bloch}$(b)$, \ref{bloch}$(c)$ and \ref{bloch}$(d)$. A second ingredient of the proof of Theorem \ref{thm_main_new} is based on the following:

\begin{prop}[Alberti-Uhlmann \cite{Alberti}]  Let $\{\sigma_1,\sigma_2\}$ and $\{\sigma_1',\sigma_2'\}$ be two sets of  qubit states. Then there exists a CPTP map $\Phi : \BH \to \BH$ connecting them i.e. $\Phi(\sigma_i)=\sigma_i'$ for $i=1,2$, if and only if
\begin{equation}
\label{eq_alberti}
 \|\sigma_1-\delta\sigma_2\|_1\geq\|\sigma_1'-\delta\sigma_2'\|_1
\end{equation}
for all $\delta>0$.
\end{prop}
Note, that the above formula can be rewritten as follows

\begin{equation}
 \|p_1 \sigma_1 - p_2\sigma_2\|_1 \geq \|p_1 \sigma_1'- p_2\sigma_2'\|_1 ,
\end{equation}
for all probability distributions $p_1+p_2=1$.

Now, the proof of Theorem \ref{thm_main_new} easily follows: condition (\ref{CP-q}) implies that $\Lambda_t$ is divisible or equivalently kernel non-decreasing \cite{PRL-2018}. Suppose now that ${\rm Im}(\Lambda_s)$ is 2-dimensional and let $\rho_1$ and $\rho_2$ be two density operators such that $\rho_1(s) = \Lambda_s(\rho_1)$ and $\rho_2(s) = \Lambda_s(\rho_2)$ span ${\rm Im}(\Lambda_s)$. Inequality (\ref{CP-q}) implies

\begin{equation}
 \|p_1 \rho_1(s) - p_2 \rho_2(s)\|_1 \geq \|p_1 \rho_1(t) - p_2\rho_2(t)\|_1 ,
\end{equation}
where $\rho_i(t) = \Lambda_t(\rho_i) = V_{t,s}(\rho_i(s))$. Alberti-Uhlmann theorem guaranties that there  exists a quantum channel $\widetilde{V}_{t,s} : \BH \to \BH$ such that $\rho_i(t) =  \widetilde{V}_{t,s}(\rho_i(s))$. Clearly, $\widetilde{V}_{t,s}$ is a CPTP extension of ${V}_{t,s} : {\rm Im}(\Lambda_s) \to \BH$.

%{\color{red} Observe that, ${\rm Im}(\Lambda_s)$ being 2-dimensional allowed us to use the Alberti-Uhlman theorem, which is a result valid only for 2 qubit states in the input and output set.}

If ${\rm Im}(\Lambda_s)$ is 1-dimensional, then $\Lambda_s(\rho) = \omega_s {\rm Tr}\rho$ for some density operator $\omega_s$. Since the map is divisible it follows that ${\rm Im}(\Lambda_t)$ is 1-dimensional for all $t>s$, and hence $V_{t,s} = \omega_t {\rm Tr}\rho$ is a CPTP projector which proves that the original qubit map $\Lambda_t$ is CP-divisible. \hfill $\Box$

It should be  emphasized that this proof requires that ${\rm Im}(\Lambda_s)$ be at most 2-dimensional otherwise one would need more than two density operators to span the image of $\Lambda_s$ and then Alberti-Uhlmann theorem is not enough to prove that there exists a universal extension for all states from the image (see also an interesting discussion of Alberti-Uhlmann theorem i \cite{Teiko}).

%\vspace{.3cm}

\section{examples}
\label{examples}

In this Section we illustrate our discussion by three examples:

\begin{enumerate}
  \item commutative and image non-increasing evolution,
  \item non-commutative but image non-increasing,
  \item non-commutative and not image non-increasing.
\end{enumerate}
Recall that if the dynamical map $\Lambda_t$ is commutative, that is, $\Lambda_{t}\Lambda_u = \Lambda_u \Lambda_t$, and diagonalizable, meaning that time independent eigenvectors $X_\alpha$ of $\Lambda_t$ and $Y_\alpha$ of the dual map $\Lambda_t^\dagger$ (so called damping basis \cite{DB})

$$   \Lambda_t[X_\alpha] = \lambda_\alpha(t) X_\alpha \ , \ \ \ \Lambda^\dagger_t[Y_\alpha] = \lambda^*_\alpha(t) Y_\alpha , $$
span the entire $\BH$, the map $\Lambda_t$ gives rise to the following spectral representation

\begin{equation}\label{}
  \Lambda_t(\rho) = \sum_\alpha \lambda_\alpha(t) X_\alpha {\rm Tr}(Y^\dagger_\alpha \rho) .
\end{equation}
Moreover, in this case if the map is divisible, that is, kernel non-decreasing, then necessarily it is image non-increasing.

\begin{ex}

A well known example of a commutative diagonalizable qubit dynamical map is generated by the following generator (it was already analyzed in \cite{PRL-2018})

\begin{equation}\label{}
  \mathcal{L}_t =    \gamma_1(t) \mathcal{L}_1 +   \gamma_2(t) \mathcal{L}_2 +  \gamma_3(t) \mathcal{L}_3   ,
\end{equation}
where $\mathcal{L}_k(\rho) = \frac 12 (\sigma_k \rho \sigma_k -\rho)$. The corresponding dynamical map reads

\begin{equation}\label{}
  \Lambda_t(\rho) = \sum_{\alpha=0}^3 p_\alpha(t) \sigma_\alpha \rho \sigma_\alpha ,
\end{equation}
with $\sigma_0 = \oper$, and

\begin{eqnarray*}
  p_0(t) &=& \frac{1}{4} [1 + \lambda_1(t) + \lambda_2(t) + \lambda_3(t)] \\
  p_1(t) &=& \frac{1}{4} [1 + \lambda_1(t) - \lambda_2(t) - \lambda_3(t)] \\
  p_2(t) &=& \frac{1}{4} [1 - \lambda_1(t) + \lambda_2(t) - \lambda_3(t)] \\
  p_3(t) &=& \frac{1}{4} [1 - \lambda_1(t) - \lambda_2(t) + \lambda_3(t)]
\end{eqnarray*}
and the corresponding eigenvalues $\lambda_\alpha(t)$ read:

$$  \lambda_i(t) = \exp( - \Gamma_j(t) - \Gamma_k(t) ) , $$
where $\{i,j,k\}$ is a permutation of $\{1,2,3\}$, and $\Gamma_k(t) = \int_0^t \gamma_k(\tau)d\tau$. The map $\Lambda_t$ is invertible if all $\Gamma_k(t)$ are finite for finite times. Now, if for example one has $\Gamma_1(t')=\infty$, then $\lambda_2(t') = \lambda_3(t'){ =  0}$ which means that the image of $\Lambda_{t'}$ is $2$-dimensional and of course it is orthogonal to the $2$-dimensional kernel:

$$  {\rm Im}(\Lambda_{t'}) = {\rm span}\{ \sigma_0,\sigma_1\} \ , \ \ {\rm Ker}(\Lambda_{t'}) = {\rm span}\{ \sigma_2,\sigma_3\} . $$
Divisibility requires that $\Gamma_1(t)=\infty$ for $t > t'$. Now, if $\Gamma_2(t)$ and $\Gamma_3(t)$ stay finite, then

$$  {\rm Im}(\Lambda_{t}) = {\rm Im}(\Lambda_{t'}) \ , \ \ {\rm Ker}(\Lambda_{t}) = {\rm Ker}(\Lambda_{t'}) , $$
for $t > t'$ { and $\Lambda_t$ is image non-increasing}. If for example $\Gamma_2(t'')=\infty$, then the image of $\Lambda_{t''}$ is 1-dimensional and it is orthogonal to the 3-dimensional kernel:

$$  {\rm Im}(\Lambda_{t''}) = {\rm span}\{ \sigma_0\} \ , \ \ {\rm Ker}(\Lambda_{t''}) = {\rm span}\{ \sigma_1,\sigma_2,\sigma_3\} . $$
Now, divisibility requires that additionally  $\Gamma_2(t)=\infty$ for $t > t''$, and as a result

$$  {\rm Im}(\Lambda_{t}) = {\rm Im}(\Lambda_{t''}) \ , \ \ {\rm Ker}(\Lambda_{t}) = {\rm Ker}(\Lambda_{t''}) , $$
for $t > t''$. {Thus in this case also $\Lambda_t$ is image non-increasing.} Finally, the map is CP-divisible iff $\gamma_1(t) \geq 0$ for $t< t'$ and $\gamma_2(t),\gamma_3(t)\geq 0$ for $t < t''$.

\end{ex}

Next example goes beyond commutative maps.

\begin{ex} Consider the following  generator \cite{Maniscalco-NJP}

\begin{equation}\label{Maniscalco_exp}
  \mathcal{L}_t =  \omega(t) \mathcal{L}_0 +  \gamma_+(t) \mathcal{L}_+ +   \gamma_-(t) \mathcal{L}_- +  \gamma_3(t) \mathcal{L}_3   ,
\end{equation}
where $\mathcal{L}_0(\rho) = -\frac{i}{2}[\sigma_z,\rho]$ is the Hamiltonian part, and

\begin{eqnarray*}
  \mathcal{L}_+(\rho) &=& \frac 12 (\sigma_+ \rho \sigma_- - \frac 12 \{ \sigma_-\sigma_+,\rho \}) , \\
  \mathcal{L}_-(\rho) &=& \frac 12 (\sigma_- \rho \sigma_+ - \frac 12 \{ \sigma_+\sigma_-,\rho \}) , \\
  \mathcal{L}_3(\rho) &=& \frac 12 (\sigma_z \rho \sigma_z - \rho)  ,
\end{eqnarray*}
with $\sigma_\pm = (\sigma_x \pm i \sigma_y)/2$. It defines a non-commutative family {in general}, that is, $\mathcal{L}_t \mathcal{L}_s \neq \mathcal{L}_s \mathcal{L}_t$. Actually, commutativity holds {if and} only if $\gamma_-(t)= k \gamma_+(t)$ for a positive constant `$k$'. {Also negative values of `$k$' would violate the CP-divisibility condition of Eq. (\ref{Maniscalco_exp}).}
The corresponding dynamical map $\Lambda_t = \mathcal{T} e^{\int_{0}^{t} \mathcal{L}_\tau d\tau}$ is given by
\begin{equation}
\label{maniscalco_evolution'}
\rho= \left(
\begin{array}{c c}
1-p & \alpha\\
\alpha^* & p
\end{array}
\right) \ \to \ \rho_t = \left(
\begin{array}{c c}
1-p(t) & \alpha(t)\\
\alpha(t)^* & p(t)
\end{array}
\right),
\end{equation}
 where
$$  p(t) = e^{-\Gamma(t)}[G(t) + p] \ , \ \ \alpha(t) = \alpha\, e^{i \Omega(t)-\Gamma(t)/2-{\Gamma_3}(t)} , $$
with  $\Gamma_3(t)=\int_0^t\gamma_3(\tau)d\tau$, $ \Omega(t)=\int_0^t\;{ 2 \;}\omega(\tau)d\tau$, and

% $$ \Gamma(t) = \frac{1}{2}\int_0^t(\gamma_+(\tau)+\gamma_-(\tau))d\tau \ , \ G(t)=\frac{1}{2}\int_0^te^{\Gamma(\tau)}\gamma_-(\tau)d\tau .
%   $$
\begin{eqnarray*}
 \Gamma(t) &=& \frac{1}{2}\int_0^t(\gamma_+(\tau)+\gamma_-(\tau))d\tau\\
 G(t)  &=&  \frac{1}{2}\int_0^te^{\Gamma(\tau)}\gamma_-(\tau)d\tau
\end{eqnarray*}

The corresponding time dependent eigenvalues of $\Lambda_t$ read: $\lambda_0(t)=1$,  $\lambda_1(t) = e^{i \Omega(t)-\Gamma(t)/2-{\Gamma}_3(t)} = \lambda_2^*(t)$, and $\lambda_3(t) = e^{-\Gamma(t)}$ (cf. Appendix \ref{appendixmaniscalco}). Finally, one finds for the time dependent eigenvectors:

$$  X_0(t)= \frac{1}{1- e^{-\Gamma(t)}}  \left( \begin{array}{c c}
1- e^{-\Gamma(t)}[G(t)+1] & 0\\
0 &  e^{-\Gamma(t)} G(t) \end{array} \right) , $$

$$   X_1 = |0\rangle \langle 1| \ , \ X_2 = |1\rangle \langle 0| \ , \ X_3 = \sigma_3 , $$
together with  $Y_0 = \openone/2$, $Y_1=X_2$, $Y_2=X_1$, and

$$  Y_3(t)= \frac{1}{1- e^{-\Gamma(t)}}  \left( \begin{array}{c c}
e^{-\Gamma(t)}G(t) & 0\\
0 &  e^{-\Gamma(t)}[ G(t)+1] -1 \end{array} \right) . $$
One easily checks that $\{X_\alpha,Y_\beta\}$ define a damping basis, that is,

$$   {\rm Tr}(X_\alpha Y^\dagger_\beta) = \delta_{\alpha\beta} . $$
The map is invertible if and only if $\Gamma(t)$ and ${\Gamma}_3(t)$ are finite for finite $t$. Now, if ${\Gamma}_3(t_1) = \infty$, then $\lambda_1(t)=\lambda_2(t)=0$ and hence $ {\rm dim}[{\rm Im}(\Lambda_{t_1})]=2$. One finds

$$  {\rm Im}(\Lambda_{t_1}) = {\rm span}\{ X_0(t),X_3\} \ , \ \ {\rm Ker}(\Lambda_{t_1}) = {\rm span}\{ X_2,X_3\} . $$
Divisibility requires that ${\Gamma}_3(t) = \infty$ for $t > t_1$, that is,

$$ {\rm Ker}(\Lambda_{t}) \supset {\rm Ker}(\Lambda_{t_1}) ={\rm span}\{ X_2,X_3\} . $$
Note , that for $t > t_1$ the image ${\rm Im}(\Lambda_{t})$ is a subset of ${\rm Im}(\Lambda_{t_1})$ since $  {\rm Im}(\Lambda_{t_1})$ spans a set of diagonal matrices. If moreover {we choose $\gamma_+$ and $\gamma_-$ in such a way that $G(t_2)$ is finite  and} ${\Gamma}(t_2) = \infty$ then also $\lambda_3(t_2)=0$ and hence $ {\rm dim}[{\rm Im}(\Lambda_{t_2})]=1$ and it is spanned by $X_0(t)$.  Again, divisibility requires that ${\Gamma}(t) = \infty$ for $t > t_2$. Note, that in this case for $t> t_2$ one has $X_0(t) = {\ket{0}\bra{0}}$ and hence

$$   \Lambda_t(\rho) = {\rm Tr}\, \rho \,{\ket{0}\bra{0}} $$
for $t \geq  t_2$, that is, all states are mapped into {the $\ket{0}\bra{0}$ state}. This proves that also this example being noncommutative gives rise to the image non increasing evolution. The evolution is CP-divisible if {$\gamma_3(t) \geq 0$ for $t\leq t_1$, and $\gamma_+(t) \geq 0$ and $\gamma_-(t)\geq 0$ for $t \leq t_2$}.

\end{ex}

{A CP-divisibility aspects of the above examples can also be studied using the results in \cite{PRL-2018}. Therefore,}
we consider now a map which is neither commutative nor image non-increasing.

\begin{ex}
Let the dynamical map be a composition of two maps:

\begin{equation}\label{}
  \Lambda_t = \mathcal{U}_t \circ \Psi_t ,
\end{equation}
where

\begin{equation}\label{}
  \mathcal{U}_t[\rho] =  U_t \rho\, U^\dagger_t  \ ; \  \ U_t = e^{-i \sigma_2 t} ,
\end{equation}
and

\begin{equation}\label{}
  \Psi_t[\rho] = [1-p(t)] \rho + p(t) \Phi[\rho] ,
\end{equation}
with

\begin{equation}\label{}
   \Phi[\rho] =    \ket{0}\bra{0}\rho\ket{0}\bra{0} + \ket{1}\bra{1}\rho\ket{1}\bra{1} ,
\end{equation}
being a totally depolarizing channel. One has $0\leq p(t)\leq 1$ with $p(0)=0$. It is clear that the map $\Lambda_t$ is CP-divisible iff the map $\Psi_t$ is CP-divisible. The map $\Lambda_t$ is invertible only if $p(t)<1$.
 Suppose now that $p(t) < 1$ for $t < t_*$ and $p(t)=1$ for $t\geq t_*$. The kernel of the map for $t \geq t_*$ is 2-dimensional

$$   {\rm Ker}(\Lambda_t) = {\rm span}\{ \sigma_1,\sigma_2\}  , $$
due to $\Phi(\sigma_1)=\Phi(\sigma_2)=0$ and hence the map is divisible.

For the image one finds

$$   {\rm Im}(\Lambda_t) = {\rm span}\{ \openone, X(t)\}  , $$
with

$$  X(t)= \left( \begin{array}{c c}
 \cos 2t & \sin 2t\\
\sin 2t  & -\cos 2t \end{array} \right) . $$
It is clear that the condition ${\rm Im}(\Lambda_t) \subset {\rm Im}(\Lambda_{t_*})$ is no longer valid and hence the map is not image non-increasing.

The corresponding propagator $V_{t,s}$ satisfies

$$   V_{t,s} = \mathcal{U}_t \circ W_{t,s} \circ \mathcal{U}_s^{-1} , $$
where $W_{t,s}$ is the propagator for the dynamical map $\Psi_t$ {i.e. $\Psi_t=W_{t,s} \circ \Psi_s$}.
Now, for $t < t_*$  the map $\Psi_t$ is invertible and one can find the corresponding time-local generator \cite{Hall}

$$  \ell_t = \dot{\Psi}_t {\Psi}^{-1}_t \ , \ \ t<t_* .$$
Using

$$  \dot{\Psi}_t[\rho] = \dot{p}(t)( \Phi(\rho) - \rho) , $$
together with

$$\Psi^{-1}_t[\rho]=\frac{1}{1-p(t)}\big(\rho-p(t)\Phi[\rho]\big)$$
one  gets

%Now, to find CP-divisibility condition we need to check if $V_{t,s}$ is CP, only for {\color{red}$s<t\le t_*$}, as for {\color{red}$t>s\ge t_*$} and $t>t_*>s$, $V_{t,s}$ and $V_{t,t_*}$ {\color{red}(using decomposition $V_{t,s}=V_{t,t_*} \circ V_{t_*,s}$)} becomes a unitary evolution. Below, we present the analysis for the the regime $t<t_*$.One has for the corresponding time-local generator

$$   \ell_t[\rho] = \frac{\dot{p}(t)}{1-p(t)} \big( \Phi [\rho] - \rho \big) . $$
Now, it is clear that the map $\Psi_t$ is CP-divisible iff
$\dot{p}(t) \geq 0$ for $t< t_*$ and $p(t) = 1 $ for $t \geq t_*$ \cite{GKLS}.
As the corresponding propagator one has

$$   W_{t,s} = \Psi_t \Psi_s^{-1}  , \ \ s < t_* , $$
and

$$   W_{t,s} =  \Phi , \ \ s \geq t_* . $$
It implies

$$   V_{t,s} = \mathcal{U}_t (\Psi_t \Psi_s^{-1} ) \mathcal{U}_s^{-1}  , \ \ s < t_* , $$
and

$$   V_{t,s} =  \mathcal{U}_t \circ \Phi \circ \mathcal{U}_s^{-1} , \ \ s \geq t_* . $$
Note, that $V_{t,t}$ is given by

$$   V_{t,t} = {\rm id} , \ \ t < t_* ,  $$
and

$$ V_{t,t} = {\mathcal{U}_{t}}\circ \Phi\circ {\mathcal{U}_{t}^{-1}}  \ , \ \ t \geq t_* .$$
It is clear that $V_{t,t}$ always defines a CPTP projector.

\end{ex}

\section{Conclusion}
\label{Conclusion}

In this paper, we discussed two main approaches to Markovianity: CP-divisibility and Information backflow, and studied conditions under which they are equivalent. This issue has recently analyzed in several papers \cite{BOGNA, PRL-2018,chakraborty} and certain classes of dynamical maps are already known for which the equivalence could be shown, namely, invertible and so called image non-increasing maps. Although the image non-increasing class includes the class of invertible maps and also several dynamical maps known in literature, it remains an open  question if the equivalence could be shown for the general case.

Here, we showed the equivalence for general qubit dynamical map (Theorem \ref{thm_main_new}).  A key element of the  proof is the fact that there are no CPTP projectors onto a qubit subspace of dimension 3 which is spanned by density operators. We also show that there can be positive projectors onto qubit subspaces of dimension 3, but only when the subspace forms an operator system. We expect this result will shed more light into the theoretical understanding of dynamical maps. In a slightly different context divisibility of qubit channels was recently addressed in \cite{Mario}.  

We also discussed the conditions for CP-divisibility and image non-increasing-ness for a number of examples of non-invertible dynamical maps. In particular, we presented an example of qubit dynamical map which is not image non-increasing to demonstrate the importance of our result.

Finally, we note that our result emphasizes the requirement of further analysis of this topic. Moreover, the question of whether the equivalence could be shown for higher dimension still remains open. This question, if proved affirmatively, will present a consistent theory of Markovianity in quantum regime.

It should be clear that our result could be immediately applied for a classical case. One has

\begin{cor} Consider a classical dynamical map represented by a family of stochastic matrices $T_t : \mathbb{R}^d \to \mathbb{R}^d$ for $t \geq 0$ and $T_{t=0} = \oper$. The map $T_t$ is called divisible if $T_t = S_{t,s} T_s$, with $S_{t,s} :  \mathbb{R}^d \to \mathbb{R}^d$ for $t \geq s$. It is called P-divisible iff $S_{t,s}$ is a stochastic matrix. Now, if $d \leq 3$, then $T_t$ is P-divisible  iff

\begin{equation}\label{}
  \frac{d}{dt} \| T_t(x_1 \mathbf{p}_1 - x_2 \mathbf{p}_2) \|_1 \leq 0 ,
\end{equation}
for arbitrary probability vectors $\mathbf{p}_1,\mathbf{p}_2 \in \mathbb{R}^d$, and probability distribution $x_1+x_2=1$. (Recall, that $L_1$-norm of $v=\{v_1,\dots,v_d\}\in\mathbb{R}^d$ is given by $\| v \|_1=\sum_{i=1}^d|v_i|$.)
\end{cor}
The proof immediately follows due to the fact that if $T_t$ is non-invertible, then its image has dimension 2 or 1.

It wold be interesting to fully clarify the problem for $d>2$ in the quantum case and $d>3$ in the classical case.

\onecolumngrid
\vspace{1cm}
%\newpage

\appendix

\section{Proof of Proposition 1}
\label{proof_proposition1}
{\it Proof 1.} We will prove the proposition by contradiction. Let a 3-dimensional subspace  spanned by qubit density operators be denoted by $\mathcal{M}$. Assume there exists a CPTP projector $\Pi_{\mathcal{M}}$ onto $\mathcal{M}$. Now consider the following arguments.

 Since $\mathcal{M}$ has dimension $3$, we must have $3$ linearly independent density matrices $\rho_1,\rho_2$ and $\rho_3$ which spans $\mathcal{M}$. As the set of all hermitian operators in $\BH$ form a real vector space, we can always find a non-zero hermitian operator $K\in\BH$ which is orthogonal to $\mathcal{M}$ i e.
 \begin{equation}
  Tr[K\rho_i]=0 ~~;~~ i=1,2,3.
  \label{innerprod}
 \end{equation}
 Let us now consider the eigenvalue decomposition of $K$ as
 \begin{equation}
  \label{eig-decomposition}
  K=\lambda_0\ket{0}\bra{0}+\lambda_1\ket{1}\bra{1},
 \end{equation}
  where $\lambda_0,\lambda_1$ are the real eigenvalues and $\{\ket{0},\ket{1}\}$ are the eigenvectors of $K$. Now, using Eq. (\ref{innerprod}) we find $\lambda_0\bra{0}\rho_i\ket{0}+\lambda_1\bra{1}\rho_i\ket{1}=0$ for $i=1,2,3$. This implies $\lambda_0\neq0,\lambda_1\neq0$, as any of them being zero will make all the $\rho_i$'s linearly dependent. Therefore, we can choose a hermitian operator $H\in\BH$ which is orthogonal to $\mathcal{M}$ and has the form
 \begin{equation}
 \label{eq_defn_H}
  H=(1/\lambda_0)K=\ket{0}\bra{0}+\lambda \ket{1}\bra{1},
 \end{equation}
where $\lambda\neq0$ is real. Note that $H\notin\mathcal{M}$ and $\{\rho_1,\rho_2,\rho_3,H\}$ forms a basis for $\BH$. Now, using Eqs. (\ref{innerprod}) and (\ref{eq_defn_H}) we get $\bra{0}\rho_i\ket{0}+\lambda\bra{1}\rho_i\ket{1}=0$ for $i=1,2,3$. Therefore we can write  all the $\rho_i$'s in the basis $\{\ket{0},\ket{1}\}$ as
\begin{align}
 \rho_1=\left(
\begin{array}{cc}
 p & x \\
 x^* & 1-p \\
\end{array}
\right)~;~
\rho_2=\left(
\begin{array}{cc}
 p & y \\
 y^* & 1-p \\
\end{array}
\right)~;~
\rho_3=\left(
\begin{array}{cc}
 p & z \\
 z^* & 1-p \\
\end{array}
\right),
\end{align}
where $x,y,z\in\mathbb{C}$ and
\begin{equation}
\label{eq_condition_p}
{ p=\frac{\lambda}{\lambda-1}.}
\end{equation}
Note that $p\neq0,p\neq1$ and $x\neq y,y\neq z, z\neq x$ as $\rho_1,\rho_2$ and $\rho_3$ are all independent. Let us define $X=\rho_1-\rho_2$ and $Y=\rho_1-\rho_3$. Note that $X,Y\in \mathcal{M}$ are independent and have the form
\begin{align}
 X=\left(
\begin{array}{cc}
 0 & u_1+i v_1 \\
 u_1-i v_1 & 0 \\
\end{array}
\right)~~~;~~~
Y=\left(
\begin{array}{cc}
 0 & u_2+i v_2 \\
 u_2-i v_2 & 0 \\
\end{array}
\right),
\end{align}
where $u_1,u_2,v_1,v_2\in\mathbb{R}$. Note that $u_1+i v_1\neq u_2+ i v_2$ as $y\neq z$. This implies either $u_1\neq u_2$ or $v_1\neq v_2$. Now since $X$ and $Y$ are independent we must also have: $(a)$ $u_1\neq0$ or $u_2\neq0$, and $(b)$ $v_1\neq0$ or $v_2\neq0$. From condition $(a)$ we find: if $u_1\neq0$,  $(u_2/u_1) X-Y=c~\sigma_y$, and, if $u_2\neq0$,  $X-(u_1/u_2)Y=c'~\sigma_y$, where $c,c'\in\mathbb{C}$ and $\sigma_y$ is the Pauli y-matrix. Similarly, from condition $(b)$ we find: if $v_1\neq0$ then $(v_2/v_1) X-Y=c''~\sigma_x$ and if $v_2\neq0$ then $X-(v_1/v_2)Y=c'''~\sigma_x$, where $c'',c'''\in\mathbb{C}$ and $\sigma_x$ is the Pauli x-matrix. This implies $\sigma_x,\sigma_y\in\mathcal{M}$ and hence $\ket{0}\bra{1},\ket{1}\bra{0}\in\mathcal{M}$. As a result we can define $Z=\rho_1-x\ket{0}\bra{1}-x^*\ket{1}\bra{0}$, where $Z\in\mathcal{M}$ and have the form
\begin{equation}
 Z=\left(
\begin{array}{cc}
 p & 0 \\
 0 & 1-p \\
\end{array}
\right).
\end{equation}
Note that $Z$ and $H$ are independent. Hence we have $\ket{0}\bra{0}=c_1H+c_2Z$ and $\ket{1}\bra{1}=c_3H+c_4Z$, where
\begin{align}
\label{eq_c_values}
 c_1=\frac{(1-p)^2}{1-2p(1-p)}~~,~~c_2=\frac{p}{1-2p(1-p)}~~,~~
 c_3=-\frac{p(1-p)}{1-2p(1-p)}~~,~~c_4=\frac{1-p}{1-2p(1-p)}.
\end{align}
Let us now consider the action of the CPTP projector $\Pi_{\mathcal{M}}$ on the following operators in $\mathcal{M}$
\begin{align}
 \Pi_{\mathcal{M}}\big[\ket{0}\bra{1}\big]=\ket{0}\bra{1}~~,~~\Pi_{\mathcal{M}}\big[\ket{1}\bra{0}\big]=\ket{1}\bra{0}~~,~~
 \Pi_{\mathcal{M}}\big[Z\big]=Z.
\end{align}
% Note, action of the map on $H$, i e. $\Pi_{\mathcal{M}}[H]$ will uniquely specify  $\Pi_\mathcal{M}$.
Let us now consider the Choi matrix  $\Gamma\big(\Pi_{\mathcal{M}}\big)$ %\cite{Choi,Jamiolkowski}
given by
\begin{align}
 \Gamma\big(\Pi_{\mathcal{M}}\big)&=\sum_{i,j=0,1}\ket{i}\bra{j}\otimes\Pi_{\mathcal{M}}\big[\ket{i}\bra{j}\big]\nonumber\\
 &=\ket{0}\bra{0}\otimes\Pi_{\mathcal{M}}\big[\ket{0}\bra{0}\big]+\ket{0}\bra{1}\otimes\ket{0}\bra{1}
 +\ket{1}\bra{0}\otimes\ket{1}\bra{0}+\ket{1}\bra{1}\otimes\Pi_{\mathcal{M}}\big[\ket{1}\bra{1}\big].
\end{align}
As $\Pi_{\mathcal{M}}$ is a CPTP map, $\Gamma\big(\Pi_{\mathcal{M}}\big)$ must be positive.  Also note that $\Pi_{\mathcal{M}}\big[\ket{0}\bra{0}\big]$ and $\Pi_{\mathcal{M}}\big[\ket{1}\bra{1}\big]$ must be density matrices. Let us choose them in the form
\begin{align}
 \Pi_{\mathcal{M}}\big[\ket{0}\bra{0}\big]=\left(
 \begin{array}{c c}
  q_1 & w_1\\
  w_1^* & 1-q_1
 \end{array}
\right)~~;~~
\Pi_{\mathcal{M}}\big[\ket{1}\bra{1}\big]=\left(
 \begin{array}{c c}
  1-q_2 & w_2\\
  w_2^* & q_2
 \end{array}
\right),
\end{align}
where $0\leq q_1,q_2\leq1$ and $w_1,w_2\in\mathbb{C}$. We can now the form of the Choi matrix
\begin{equation}
 \Gamma\big(\Pi_{\mathcal{M}}\big)=\left(
 \begin{array}{c c c c}
  q_1 & w_1 & 0 & 1\\
  w_1^* & 1-q_1 & 0 & 0\\
  0 & 0 & 1-q_2 & w_2\\
  1 & 0 & w_2^* & q_2
 \end{array}
\right).
\end{equation}
It can be easily seen from the above form that since $\Gamma\big(\Pi_{\mathcal{M}}\big)$ is positive we must have $q_1q_2-1\geq0$, which is only possible when $q_1=q_2=1$. As a result, preserving positivity of $\Pi_{\mathcal{M}}\big[\ket{0}\bra{0}\big]$ and $\Pi_{\mathcal{M}}\big[\ket{1}\bra{1}\big]$ we  get $w_1=w_2=0$. This implies
$\Pi_{\mathcal{M}}\big[\ket{0}\bra{0}\big]=\ket{0}\bra{0}$ and $\Pi_{\mathcal{M}}\big[\ket{1}\bra{1}\big]=\ket{1}\bra{1}$. This in turn implies $\Pi_{\mathcal{M}}[H]=H$. As a result, $\text{Im}(\Pi_{\mathcal{M}})\not\subset\mathcal{M}$, which is a contradiction.\\

{\it Proof 2.} The proposition also follows from a result in \cite{QUBIT}. It was shown in Theorem $4.9$ of the paper, that if any qubit channel (CPTP map) has more than $2$ pure states in its output it must contain all pure states in its output. From our analysis in the above proof it can be easily seen that the infinite family of pure states
  \begin{equation}
\Big\{\ket{\psi_{\theta}}=\sqrt{p}\ket{0}+\sqrt{1-p}e^{i\theta}\ket{1};\theta\in\mathbb{R}\Big\}\subset\mathcal{M},
\end{equation}
appears in the output of $\Pi_{\mathcal{M}}$, but the states $\ket{0}$ and $\ket{1}$ does not. Hence $\Pi_{\mathcal{M}}$ cannot be a CPTP map.

\section{Are there PTP projectors on 3 dimensional subspaces spanned by qubit states?}
\label{positive-projector}
We consider a qubit space i e. $\mathcal{H}=\mathbb{C}^2$. Let us consider a PTP projector $\pi_{\mathcal{M}}$ onto $\mathcal{M}$, where $\mathcal{M}$ is as defined in Appendix \ref{proof_proposition1}. Therefore, $\pi_{\mathcal{M}}$ must have the following properties:
\begin{enumerate}[label={(\alph*)}]
 \item $\text{Im}(\pi_{\mathcal{M}})\subset\mathcal{M}$.
 \item $\pi_{\mathcal{M}}[X]=X$ if $X\in \mathcal{M}$.
\end{enumerate}
Following the same analysis as in Appendix \ref{proof_proposition1}, we find
\begin{align}
 \pi_{\mathcal{M}}\big[\ket{0}\bra{1}\big]=\ket{0}\bra{1}~,~\pi_{\mathcal{M}}\big[\ket{1}\bra{0}\big]=\ket{1}\bra{0}~,~
 \pi_{\mathcal{M}}\big[Z\big]=Z.
\end{align}
Now consider the action of $\pi_{\mathcal{M}}$ on $H$, using the form given in Eq. (\ref{eq_defn_H}). As $\pi_{\mathcal{M}}$ is TP, without loss of generality, we get
\begin{align}
\label{eq_defn_H1}
 \pi_{\mathcal{M}}[H]~=~ (1+\lambda)Z+s\ket{0}\bra{1}+s^*\ket{1}\bra{0}
 ~=~\frac{1-2p}{1-p}Z+s\ket{0}\bra{1}+s^*\ket{1}\bra{0},
\end{align}
where $s\in\mathbb{C}$. Here we have used the relation between $p$ and $\lambda$, as given in Eq. (\ref{eq_condition_p}).

Now consider any state $\rho\in\mathcal{P}_+(\mathcal{H})$ having the following form in the $\{\ket{0},\ket{1}\}$ basis.
\begin{equation}
\label{eq_defn_rho}
 \rho=\left(
 \begin{array}{c c}
  q & r\\
  r^* & 1-q
 \end{array}
\right),
\end{equation}
where $0\leq q\leq1, r\in\mathbb{C}$ and $|r|^2\leq q(1-q)$. Note that, this is the most general form of a qubit state. Also note that, the maximum value that $|r|$ can take, is $1/2$ which is when $\rho=\ket{\pm}\bra{\pm}$, where $\ket{\pm}=\frac{1}{\sqrt{2}}(\ket{0}\pm\ket{1})$. Now, consider the action of $\pi_{\mathcal{M}}$ on $\rho$ :
\begin{align}
 \pi_{\mathcal{M}}[\rho]&=q\;\pi_{\mathcal{M}}\big[\ket{0}\bra{0}\big]+(1-q)\;\pi_{\mathcal{M}}\big[\ket{1}\bra{1}\big]+r\ket{0}\bra{1}+r^*\ket{1}\bra{0}\nonumber\\
 &=q\;\pi_{\mathcal{M}}\big[c_1H+c_2Z\big]+(1-q)\;\pi_{\mathcal{M}}\big[c_3H+c_4Z\big]+r\ket{0}\bra{1}+r^*\ket{1}\bra{0}\nonumber\\
 &=\big[q\;c_1+(1-q)c_3\big]\pi_{\mathcal{M}}\big[H\big]+\big[q\;c_2+(1-q)c_4\big]Z+r\ket{0}\bra{1}+r^*\ket{1}\bra{0}\nonumber\\
 &=\big[q\;c_1+(1-q)c_3\big]\frac{1-2p}{1-p}Z+\big[q\;c_2+(1-q)c_4\big]Z+(r+s)\ket{0}\bra{1}+(r^*+s^*)\ket{1}\bra{0}\nonumber\\
 &=\Big[q\Big(c_1\frac{1-2p}{1-p}+c_2\Big)+(1-q)\Big(c_3\frac{1-2p}{1-p}+c_4\Big)\Big]Z+(r+s)\ket{0}\bra{1}+(r^*+s^*)\ket{1}\bra{0}.
\end{align}
Now putting the values of $c_1$ and $c_2$, as given in Eq. (\ref{eq_c_values}), we find
$c_1\frac{1-2p}{1-p}+c_2=c_3\frac{1-2p}{1-p}+c_4~=1$. As a result, we find
\begin{equation}
 \pi_{\mathcal{M}}[\rho]=\left(
 \begin{array}{c c}
  p & r+s\\
  r^*+s^* & 1-p
 \end{array}
\right)
\end{equation}
In the above form if $\pi_{\mathcal{M}}[\rho]$ is positive, we  must have
\begin{equation}
\label{condition on r,s}
 |r+s|^2\leq p(1-p)\leq 1/4,
\end{equation}
where we have used the fact $p(1-p)\leq 1/4$ and the equality is reached only when $p=1/2$. Note in Eq. (\ref{eq_defn_H1}) that $s$ can take any complex value. Let $s=s_1+is_2$ ($s_1,s_2\in\mathbb{R}$). If we consider either of the real or imaginary parts of $s$ to be non-zero, we can always choose $\rho$ in Eq. (\ref{eq_defn_rho}) to have $q=1/2$ and $r=\frac{s_1}{2|s_1|}$, and as a result we get $|r+s|^2=(\frac{1}{2}+|s_1|)^2+|s_2|^2 > \frac{1}{4}$, which contradicts condition (\ref{condition on r,s}). Therefore, we must have $s=0$. Now, if we choose $\rho$ in Eq. (\ref{eq_defn_rho}) to have $q=1/2$ and $r=1/2$, the inequality in Eq. (\ref{condition on r,s}) will be satisfied only when $p=1/2$. Hence we conclude $\pi_{\mathcal{M}}$ is a PTP map only when $p=1/2$. In that case, $\mathcal{M}={\rm span}\big\{\frac{\openone}{2},\frac{\openone+
\sigma_1}{2},\frac{\openone+\sigma_2}{2}\big\}$ and the PTP projector has the action: $\openone\rightarrow\openone~,~\sigma_1\rightarrow\sigma_1~,~\sigma_2\rightarrow\sigma_2~,~\sigma_3\rightarrow0$. In other words, a PTP projector on a $3$ dimensional qubit subspace spanned by density matrices will only exist, when the subspace is an {\it Operator System} \cite{bhatia2009positive}.

\section{ Calculations for Example 2}
\label{appendixmaniscalco}
The time evolved state in Maniscalco's paper \cite{Maniscalco-NJP} is given by
\begin{equation}
\label{maniscalco_evolution}
\rho(t)=\Lambda_t[\rho(0)]=\left(
\begin{array}{c c}
1-P_1(t) & \alpha(t)\\
\alpha(t)^* & P_1(t)
\end{array}
\right),
\end{equation}
 where
 \begin{align}
 P_1(t)&=e^{-\Gamma(t)}[G(t)+P_1(0)],\label{P1(t)}\\
 \alpha(t)&=\alpha(0)e^{i \Omega(t)-\Gamma(t)/2-\Gamma_3(t)}.\label{alpha(t)}
 \end{align}
 Note that $G(t),\Omega(t),\Gamma(t)$ and $\Gamma_3(t)$ are as defined in Example 2.
 Now consider the states

  \begingroup
\setlength{\tabcolsep}{100pt} % Default value: 6pt
\renewcommand{\arraystretch}{0.5} % Default value: 1
\begin{table}
\label{table1}
\caption{Eigenvalues and eigenvectors of $\mathcal{N}_t$}
\begin{ruledtabular}
\begin{tabular}{c c}
   {\bf Eigenvalues} & {\bf Eigenvectors} \\
  \hline
  &\\
  $1$ & $\Big\{\frac{1-e^{-\Gamma(t)} (G(t)+1)}{e^{-\Gamma(t)} G(t)},0,0,1\Big\}$ \\
  &\\
  $e^{-\Gamma(t)}$ & $\{-1,0,0,1\}$ \\
  &\\
  $e^{i \Omega(t)-\Gamma(t)/2-\Gamma_3(t)}$ & $\{0,1,0,0\}$ \\
  &\\
  $e^{-i \Omega(t)-\Gamma(t)/2-\Gamma_3(t)}$ & $\{0,0,1,0\}$
 \end{tabular}
\end{ruledtabular}
\label{table1}
\end{table}
\endgroup

 \begin{align}
 \ket{0}\bra{0}=\frac{1}{2}\left(
 \begin{array}{c c}
 1 & 0\\
 0 & 0
 \end{array}
 \right)~~;~~
 \ket{1}\bra{1}=\frac{1}{2}\left(
 \begin{array}{c c}
 0 & 0\\
 0 & 1
 \end{array}
 \right)~~;~~
 \ket{+}\bra{+}=\frac{1}{2}\left(
 \begin{array}{c c}
 1 & 1\\
 1 & 1
 \end{array}
 \right)~~;~~
 \ket{+_y}\bra{+_y}=\frac{1}{2}\left(
 \begin{array}{c c}
 1 & i\\
 -i & 1
 \end{array}
 \right).
 \end{align}
 Using these states, we can easily find
 \begin{align}
 \ket{0}\bra{1}&=\ket{+}\bra{+}-i\ket{+_y}\bra{+_y}
 -\frac{1-i}{2}\Big(\ket{0}\bra{0}+\ket{1}\bra{1}\Big),\\
 \ket{1}\bra{0}&=\ket{+}\bra{+}+i\ket{+_y}\bra{+_y}
 -\frac{1+i}{2}\Big(\ket{0}\bra{0}+\ket{1}\bra{1}\Big).
 \end{align}
 Now, using  Eq. (\ref{maniscalco_evolution}) we get the time evolved states,
 \begin{align}
 \Lambda_t\big[\ket{0}\bra{0}\big]&=\left(
 \begin{array}{c c}
 1 - e^{-\Gamma(t)} G(t) & 0\\
 0 & e^{-\Gamma(t)} G(t)
 \end{array}
 \right);\\
 \Lambda_t\big[\ket{1}\bra{1}\big]&=\left(
 \begin{array}{c c}
 1 - e^{-\Gamma(t)} (G(t)+1) & 0\\
 0 & e^{-\Gamma(t)} (G(t)+1)
 \end{array}
 \right);\\
 \Lambda_t\big[\ket{+}\bra{+}\big]&=\left(
 \begin{array}{c c}
 1 - e^{-\Gamma(t)} (G(t)+\frac{1}{2}) & \frac{1}{2} e^{i \Omega(t)-\Gamma(t)/2-\Gamma_3(t)}\\
 \frac{1}{2} e^{-i \Omega(t)-\Gamma(t)/2-\Gamma_3(t)} & e^{-\Gamma(t)} (G(t)+\frac{1}{2})
 \end{array}
 \right);\\
 \Lambda_t\big[\ket{+_y}\bra{+_y}\big]&=\left(
 \begin{array}{c c}
 1 - e^{-\Gamma(t)} (G(t)+\frac{1}{2}) & \frac{i}{2} e^{i \Omega(t)-\Gamma(t)/2-\Gamma_3(t)}\\
 \frac{-i }{2} e^{-i \Omega(t)-\Gamma(t)/2-\Gamma_3(t)} & e^{-\Gamma(t)} (G(t)+\frac{1}{2})
 \end{array}
 \right).
 \end{align}
 As a result, we can also find the action of $\Lambda_t$ on operators, which are not density matrices.
 \begin{align}
 \Lambda_t\big[\ket{0}\bra{1}\big]&=\left(
 \begin{array}{c c}
 0 & e^{i \Omega(t)-\Gamma(t)/2-\Gamma_3(t)}\\
 0 & 0
 \end{array}
 \right);\\
 \Lambda_t\big[\ket{1}\bra{0}\big]&=\left(
 \begin{array}{c c}
 0 & 0\\
 e^{-i \Omega(t)-\Gamma(t)/2-\Gamma_3(t)} & 0
 \end{array}
 \right)
 \end{align}
 Now consider the {\it operator-vector correspondence} described in the following way:
 the vector correspondent of an operator $A=\sum_{i,j}\ket{i}\bra{j}$ is defined as the vector $vec(A)=\sum_{i,j}\ket{i}\ket{j}$. Therefore, using this notation we define $\mathcal{N}_t$ in the following way
 \begin{equation}
vec\big(\rho(t)\big)= \mathcal{N}_t\Big(vec\big(\rho(0)\big)\Big).
 \end{equation}
 Note that $\mathcal{N}_t$ is a $4\times 4$ matrix of the following form,
 \begin{equation}
 \label{matrixformap}
 \mathcal{N}_t=\left(
 \begin{array}{c c c c}
 1 - e^{-\Gamma(t)} G(t) & 0 & 0 & 1 - e^{-\Gamma(t)} (G(t)+1)\\
 0 & e^{i \Omega(t)-\Gamma(t)/2-\Gamma_3(t)} & 0 & 0\\
 0 & 0 & e^{-i \Omega(t)-\Gamma(t)/2-\Gamma_3(t)} & 0\\
 e^{-\Gamma(t)} G(t) & 0 & 0 &  e^{-\Gamma(t)} (G(t)+1)
 \end{array}
 \right).
 \end{equation}
 $\mathcal{N}_t$ has the same eigenvalues as the qubit map $\Lambda_t$. We found the eigenvalues and eigenvectors of $\mathcal{N}_t$ to be of the following form given in Table \ref{table1}.

\section*{Acknowledgements}  DC was partially supported by the National Science Centre project 2015/17/B/ST2/02026. SC is thankful to Suchetana Goswami for useful discussions on outlining and preparing the figure. We thank \'Angel Rivas for valuable comments.

\twocolumngrid

\vspace{.4cm}

%\acknowledgements

%\bibliography{Sagnik-letter.bib}

\begin{thebibliography}{1} \bibliographystyle{plain}

\bibitem{open1} H.-P. Breuer and F. Petruccione,
{\em The Theory of Open Quantum Systems} (Oxford Univ. Press,
Oxford, 2007).

\bibitem{open2} U. Weiss, {\em Quantum Dissipative Systems}, (World Scientific,
Singapore, 2000).

\bibitem{open3} A. Rivas and S. F. Huelga, {\it Open Quantum Systems. An
Introduction} (Springer, Heidelberg, 2011).

\bibitem{NM1} \'A. Rivas, S. F. Huelga, and M. B. Plenio, Rep. Prog. Phys.
{\bf 77}, 094001 (2014).



\bibitem{exp} B.-H. Liu, et al, % Li Li, Y.-F. Huang, C.-F. Li, G.-C. Guo, E.-M. Laine, H.-P. Breuer, and J. Piilo,
Nature Physics {\bf 7}, 931 (2011).

\bibitem{exp2} N. K. Bernardes, et al, % A. Cuevas, A. Orieux, C. H. Monken, P. Mataloni, F. Sciarrino, and M. F. Santos,
Scien. Rep. {\bf 5}, 17520 (2015).

\bibitem{exp3} J. Jin et al, 	Phys. Rev. A {\bf 91}, 012122 (2015)



\bibitem{NM2} H.-P. Breuer, E.-M. Laine, J. Piilo, and B. Vacchini,
Rev. Mod. Phys. {\bf 88}, 021002 (2016).

\bibitem{NM3} I. de Vega and D. Alonso, Rev. Mod. Phys. {\bf 89},  015001  (2017).


\bibitem{NM4}  Li Li, M. J. W. Hall, and H. M. Wiseman, {\em Concepts of quantum non-Markovianity: a hierarchy}, Phys. Rep. (2018) doi.org/10.1016/j.physrep.2018.07.001


\bibitem{RHP} \'A. Rivas, S.F. Huelga, and M.B. Plenio, Phys. Rev. Lett. {\bf 105}, 050403
(2010).

\bibitem{BLP} H.-P. Breuer, E.-M. Laine, and J. Piilo, Phys. Rev. Lett. {\bf 103},
210401 (2009).

\bibitem{footnote1} Actually, one may introduce the whole hierarchy of $k$-divisibility: $\Lambda_t$ is $k$-divisible if the map $V_{t,s}$ is $k$-positive on the entire $\BH$. See D. Chru\'sci\'nski and S. Maniscalco, Phys. Rev. Lett. {\bf 112}, 120404 (2014).



\bibitem{NC00} M. A. Nielsen and I. L. Chuang {\it Quantum Computation and Quantum Information} (Cambridge University Press, Cambridge, 2000).




\bibitem{Angel} D. Chru\'sci\'nski, A. Kossakowski, and \'A. Rivas,  Phys. Rev. A {\bf 83}, 052128 (2011).

%\bibitem{HEL} C. W. Helstrom, {\it Quantum Detection and Estimation Theory} (Academic Press, New York, 1976).

\bibitem{BOGNA} B. Bylicka, M. Johansson, and A. Ac\'in, Phys. Rev. Lett. {\bf 118}, 120501 (2017).

\bibitem{PRL-2018} D. Chru\'sci\'nski, \'A. Rivas, and E. St{\o}rmer, Phys. Rev. Lett. {\bf 121}, 080407 (2018).

\bibitem{chakraborty} S. Chakraborty, Phys. Rev. A {\bf 97}, 032130 (2018).

\bibitem{datta} F. Buscemi and N. Datta, Phys. Rev. A {\bf 93}, 012101 (2016).


\bibitem{QUBIT} D. Braun, O. Giraud, I. Nechita, C. Pellegrini, and M. Z\'nidaric, J. Phys. A {\bf  47}, 135302 (2014).

\bibitem{Alberti} P. Alberti and A. Uhlmann, Rep. Math. Phys. {\bf 18}, 163 (1980).

\bibitem{Maniscalco-NJP} J. Teittinen, H. Lyyra, B. Sokolov, and S. Maniscalco, New Journal of Physics {\bf 20}, 073012 (2018).

\bibitem{GKLS} V. Gorini, A. Kossakowski, E.~C.~G. Sudarshan, J. Math. Phys. \textbf{17}, 821 (1976); G. Lindblad, Commun. Math. Phys. \textbf{48}, 119 (1976).


%\bibitem{SM} See the supplementary material document.

%\bibitem{Kossak} A. Kossakowski, Rep. Math. Phys. \textbf{3}, 247 (1972); Bull. Acad. Pol. Sci. Math. Ser. Math. Astron. {\bf 20}, 1021 (1972).
%
%\bibitem{Ruskai} M. B. Ruskai, Rev. Math. Phys. {\bf 6}, 1147 (1994).


\bibitem{Paulsen} V. Paulsen, {\em Completely Bounded Maps and Operator Algebras}, (Cambridge University Press, 2003).

\bibitem{Erling} E. St{\o}rmer, {\em Positive linear maps of operator algebras}, (Springer Monographs in Mathematics, 2013).

%\bibitem{Arveson} W. B. Arveson,  Acta Math. {\bf 123}, 141 (1969).

%\bibitem{Jencova} A. Jencova, J. Math. Phys. {\bf 53}, 012201 (2012).

\bibitem{Teiko} T. Heinosaari, M. A. Jivulescu, D. Reeb, and M. M. Wolf, J. Math. Phys. {\bf 53}, 102208 (2012).


\bibitem{DB} H.-J.Briegel and B.-G. Englert, Phys. Rev. A {\bf 47}, 3311 (1993).

\bibitem{Hall} Conditions to write a time-local master equation for a non-invertible dynamical map have been considered in E. Andersson, J. D. Cresser, and M. J. W. Hall, J. Mod. Opt. \textbf{54}, 1695 (2007).
%we have taken the notation $\mathds{1}\otimes\Lambda\equiv\mathds{1}_d\otimes\Lambda$ with d=\dim(\mathcal{H})throughout the text.


%\bibitem{Choi} M.-D. Choi, Linear Algebra and its Applications {\bf 10}, 285 (1975).

%\bibitem{Jamiolkowski} A. Jamiolkowski, Reports on Math. Phys. {\bf 3}, 275 (1972).

\bibitem{bhatia2009positive} R. Bhatia, Positive definite matrices, Vol. {\bf 24} (Princeton university press, 2009).

\bibitem{Mario} D. Davalos, M. Ziman, and C. Pineda, {\em Divisibility of qubit channels and dynamical maps}, arXiv:1812.11437. 

\end{thebibliography}
%\bibliographystyle{ieeetr}
%\bibliographystyle{apsrev}

\end{document}